\documentclass[epj]{svjour}
\usepackage{epsfig}
\usepackage{psfig,graphicx,float,pst-all,amssymb}

\newcommand{\be}{\begin{equation}}
\newcommand{\ee}{\end{equation}}
\newcommand{\bc}{\begin{center}}
\newcommand{\ec}{\end{center}}

\begin{document}
\title{Soft triaxial rotor in the vicinity of $\gamma=\pi/6$ and its 
extensions}

\author{L. Fortunato, S. De Baerdemacker and K.Heyde}

\institute{Vakgroep Subatomaire en Stralingfysica, University of Gent 
(Belgium), Proeftuinstraat, 86 B-9000, Gent (Belgium)
\\ \email{fortunat@pd.infn.it} }

\abstract{The collective Bohr hamiltonian is solved for the soft
triaxial rotor around $\gamma_0=\pi/6$ with a displaced harmonic oscillator
potential in $\gamma$ and a Kratzer-like potential in $\beta$. The properties
of the spectrum are outlined and a generalization for the more general
triaxial case with $0<\gamma<\pi/6$ is proposed. 
}

\maketitle
\noindent
Analytic or approximated solutions of the Bohr collective model
may be given for a variety of different model potentials. The functional
dependence of this potential on the deformation ($\beta$) and asymmetry 
($\gamma$) variables determines the properties of the spectrum and 
eigenfunctions. These solutions are not limited to rigid cases (where
either one or two of the variables are constrained to take a fixed value),
but may be found in the case of soft potentials too (here the potential 
function is represented by a well and it is associated with extended 
wave functions). 
A soft solution represents a more physical case than a rigid one and
a mathematical benchmark for our understanding of collective states in 
nuclear spectroscopy. Recently
Iachello introduced new solutions based on the infinite square well potential,
 named E(5), X(5) and Y(5), to describe
the critical point of shape phase transitions \cite{Iac}. These solutions
have initiated on one side intense and successful efforts aimed at the 
identification of the predicted patterns (nuclear spectra
and electromagnetic properties) in experimentally observed spectroscopic data. 
On the other side a number of theoretical studies have explored new analytic
solutions in various cases, from $\gamma-$unstable to axial rotor \cite{FV}.

\noindent
A solution of the stationary Schr\"odinger equation
\be
H_B \Psi(\beta,\gamma, \theta_i)=E  \Psi(\beta,\gamma, \theta_i)\,,
\label{un}
\ee
for the Bohr collective hamiltonian, $H_B= T_\beta+T_\gamma+T_{rot}+
V(\beta,\gamma) $ may be achieved
for the $\beta-$soft, $\gamma-$soft triaxial rotor
making use of a harmonic potential in $\gamma$ and  Coulomb-like and
Kratzer-like potentials in $\beta$:
\be
V(\beta, \gamma)= V_1(\beta)+{V_2(\gamma)\over \beta^2}\,, 
\label{du}
\ee
with
\be
V_1(\beta)=  -{A\over \beta}+
{B\over \beta^2}\,, \quad \quad V_2(\gamma)=C
(\gamma-\gamma_0)^2 \,.
\label{tr}
\ee
Unimportant multiplicative factors have been omitted here for simplicity. 
The Schr\"odinger equation above, (\ref{un}), with the choice (\ref{du}),
is separable and can be solved in the vicinity of $\gamma_0=\pi/6$, thus 
providing a paradigm for the spectrum of soft triaxial rotors.

\begin{figure}[t]
\begin{center}
\begin{picture}(180,180)(0,0)
\psset{unit=1pt}
\rput(90,90){\epsfig{file=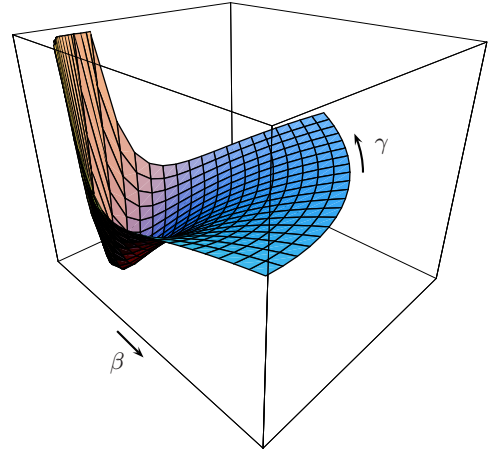,width=0.35\textwidth}}
\psline{->}(40,50)(50,40)\rput(40,38){\small $\beta$}
\psarc{->}(98,110){35}{0}{25} \rput(140,120){\small $\gamma$}
\end{picture}
\end{center}
\caption{Polar plot of the potential $V(\beta,\gamma)$ discussed in the text
with minimum in $\gamma_0=\pi/6$ and $\beta=0.2$.}
\end{figure}
\noindent
It has been shown in \cite{LF} that the $\gamma-$angular part in the 
present case
gives rise to a straightforward extension of the rigid triaxial rotor
energy, also called Meyer-ter-Vehn formula \cite{MTV}, in which now an 
additive harmonic term appears, namely
\be
\omega_{L,R,n_\gamma} = \sqrt{C}(2n_\gamma+1) + L(L+1)-{3\over 4}R^2
\label{qu}
\ee
where $R$ is the quantum number associated with the projection of 
the angular momentum on the intrinsic 1-axis (that is a good quantum number 
for the $\gamma=\pi/6$ rotor \cite{MTV}).

\begin{figure}[!t]
\epsfig{file=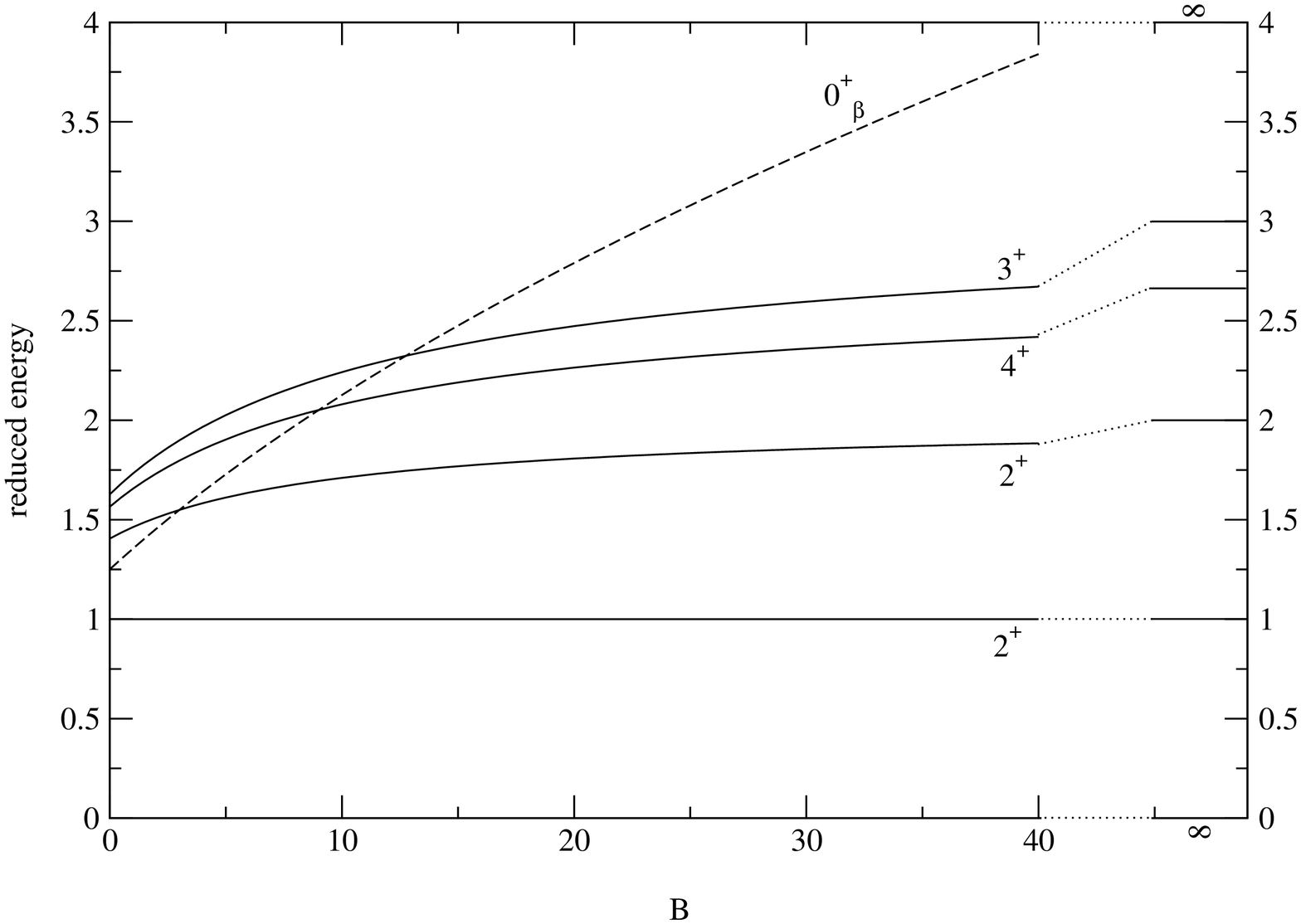, bbllx=0, bblly=0, bburx=732, bbury=570,
width=0.48\textwidth}
\caption{Reduced energies of the lowest state of the $\beta-$band
(dashed line) and of a few lowest states of the ground state band
(solid lines) as a function of $B$. The limits for the energy levels
when $B\rightarrow \infty$, that correspond to the rigid triaxial rotor
energies, are reported in the right side. Here we fixed $C=1$. From \cite{LF}.}
\label{fig2}
\end{figure}

\noindent
The solution of the equation in $\beta$ depends on the particular 
choice of the $\beta-$potential and may results instead in a non-trivial 
expression for the energy spectrum. Using the Kratzer-like potential
we obtain:
\be
\epsilon(n_\gamma,n_\beta,L,R)={A^2/4 \over \Bigl(\sqrt{9/4+B+
\omega_{L,R,n_\gamma}} +1/2+n_\beta \Bigr)^2}\,.
\ee
The negative anharmonicities of the energy levels with respect to a
simple rigid model are in qualitative agreement with general trends as
observed in experimental data. This model is more general than the 
Davydov (rigid) model \cite{Dav}: in fact the rigid model is recovered
when the potential well becomes very narrow (that is when $B\rightarrow 
\infty$) as can be seen in the right side of fig. \ref{fig2}.

\noindent
Here we present the expression of the spectrum in another well-known
solvable case: the Davidson potential, $A_D \beta^2+B_D/\beta^2$ , 
discussed in \cite{RoBa,LSK} and references therein. The spectrum may again be
found in an analytical way. We obtain
\be
\epsilon_D(n_\gamma,n_\beta,L,R)=\sqrt{A_D} \Bigl( 2n_\beta + 
\tau_{L,R,n_\gamma}+5/2\Bigr)\,,
\ee
where $\tau$ is found from $(\tau+1)(\tau+2)=B_D-\omega_{L,K,n_\gamma}$.

\noindent
Recently it has become possible to extend these results to soft triaxial
rotors with a harmonic potential (as in eq. (\ref{tr}) on the right) 
centered around any asymmetry in the sector $0<\gamma_0 <\pi/3$ by means of a
group theoretical approach based on the su(1,1) algebra \cite{LSK}.
Here the labeling is more difficult since neither $K$ nor $R$ (quantum
numbers associated with the projections of the third component of the angular 
momentum on the 3rd and 1st intrinsic axis respectively) are good quantum 
numbers, but a classification of the states is still possible on the basis
of the remaining quantum numbers. 

\noindent
Retaining the same procedure used in the $\gamma_0=\pi/6$ case for the
separation of variables we are faced with the problem of solving the 
equation in $\gamma$ that contains a rather complicated rotational kinetic
term (Here a simplification like the one used in \cite{Iac} (2nd paper) 
or \cite{LF} may not be adopted).
The components of the moment of inertia that occur in that term 
are simplified here, neglecting fluctuations in the $\gamma-$variable, in
the following way
\be
A_\kappa={1\over 4 \sin^2(\gamma-2\pi\kappa/3)} \longrightarrow 
{1\over 4 \sin^2(\gamma_0-2\pi\kappa/3)}\,.
\ee
The equation in $\gamma$ is then transformed in a set of coupled differential
equations by expanding the (general triaxial) wave functions in a basis of 
rotational (axial) wave functions. Introducing also some 
standard trigonometric approximations it is possible to 
define a realization of the algebra su(1,1)
in terms of differential operators with which, for each $L$, we can reduce the 
secular problem to an algebraic equation. When the algebraic equation has a 
low order it can be solved analytically, while for higher orders one can
always get a numerical (accurate) solution. The results have the same 
structure of eq. (\ref{qu}): the 'rotational part', that coincides in every
detail with the well-known solution of the rigid model, is accompanied by an
additive harmonic term, that takes into account the $\gamma$ quanta.

\noindent
Once $\omega$ is obtained, it must be used in the differential equation in 
$\beta$, that may be solved in standard ways, depending again on the 
$\beta-$potential. This extension automatically generates the particular
results obtained above when $\gamma_0=\pi/6$.

\noindent
This model contains in total 3 parameters (2 from the $\beta$ and $\gamma$ 
potentials, $B$ and $C$, and one from the moments of inertia, $A_3$, 
or alternatively $\gamma_0$) and may provide
a simple model for the interpretation of collective spectra of
a large number of nuclei that do not posses axial symmetry. The dependence
of the reduced spectrum on the three parameters is however non-linear and
at present we have only applied the model to spectroscopic data in a 
preliminary way. A more complete description of the problem, of the
methodology used and applications will soon be presented \cite{LSK}.

~\\
\noindent
We acknowledge financial support from ``FWO-Vlaanderen'' and 
``Universiteit Gent'' (Belgium).

\end{document}